\documentclass[
reprint,
superscriptaddress,
frontmatterverbose, 
amsmath,amssymb,
aps,
prl,
]{revtex4-2}

\usepackage[rightcaption]{sidecap}

\usepackage{graphicx}
\usepackage{dcolumn}
\usepackage{bm}
\usepackage{amsmath}
\usepackage{amssymb}
\usepackage{CJK}
\usepackage{keyval}
\usepackage{color}
\usepackage{txfonts}
\usepackage{verbatim}
\usepackage{lineno} 

\usepackage{booktabs}

\makeatletter

\newcommand{\Rmnum}[1]{\expandafter\@slowromancap\romannumeral #1@}
\makeatother

\begin{document}

\preprint{APS/123-QED}

\title{A Disconnected Superconducting Regime at the Parent Limit of Infinite-Layer Nickelates}


\author{Chihao Li}
\thanks{These authors contributed equally to this work.}
\affiliation{State Key Laboratory of Surface Physics, and Department of Physics, Fudan University, Shanghai 200438, China}
\affiliation{Laboratory of Advanced Materials, Fudan University, Shanghai 200438, China}

\author{Yutong Chen}
\thanks{These authors contributed equally to this work.}
\affiliation{State Key Laboratory of Surface Physics, and Department of Physics, Fudan University, Shanghai 200438, China}
\affiliation{Laboratory of Advanced Materials, Fudan University, Shanghai 200438, China}

\author{Yaolong Bian}
\thanks{These authors contributed equally to this work.}
\affiliation{Anhui Key Laboratory of Low-Energy Quantum Materials and Devices, High Magnetic Field Laboratory (CHMFL), Hefei Institutes of Physical Science, Chinese Academy of Sciences, Hefei, China}
\affiliation{Science Island Branch of Graduate School, University of Science and Technology of China, Hefei, China}

\author{Yihao Zhang}
\affiliation{State Key Laboratory of Surface Physics, and Department of Physics, Fudan University, Shanghai 200438, China}
\affiliation{Laboratory of Advanced Materials, Fudan University, Shanghai 200438, China}

\author{Jiahao Ye}
\affiliation{State Key Laboratory of Surface Physics, and Department of Physics, Fudan University, Shanghai 200438, China}
\affiliation{Laboratory of Advanced Materials, Fudan University, Shanghai 200438, China}

\author{Zhitong An}
\affiliation{State Key Laboratory of Surface Physics, and Department of Physics, Fudan University, Shanghai 200438, China}
\affiliation{Laboratory of Advanced Materials, Fudan University, Shanghai 200438, China}

\author{Xingtian Sun}
\affiliation{State Key Laboratory of Surface Physics, and Department of Physics, Fudan University, Shanghai 200438, China}
\affiliation{Laboratory of Advanced Materials, Fudan University, Shanghai 200438, China}

\author{Yu Fan}
\affiliation{State Key Laboratory of Surface Physics, and Department of Physics, Fudan University, Shanghai 200438, China}
\affiliation{Laboratory of Advanced Materials, Fudan University, Shanghai 200438, China}

\author{Zhihui Chen}
\affiliation{State Key Laboratory of Surface Physics, and Department of Physics, Fudan University, Shanghai 200438, China}
\affiliation{Laboratory of Advanced Materials, Fudan University, Shanghai 200438, China}

\author{Zhanze Wang}
\affiliation{State Key Laboratory of Surface Physics, and Department of Physics, Fudan University, Shanghai 200438, China}
\affiliation{Laboratory of Advanced Materials, Fudan University, Shanghai 200438, China}

\author{Jinglei Zhang}
\email{zhangjinglei@hmfl.ac.cn}
\affiliation{Anhui Key Laboratory of Low-Energy Quantum Materials and Devices, High Magnetic Field Laboratory (CHMFL), Hefei Institutes of Physical Science, Chinese Academy of Sciences, Hefei, China}

\author{Haichao Xu}
\email{xuhaichao@fudan.edu.cn}
\affiliation{State Key Laboratory of Surface Physics, and Department of Physics, Fudan University, Shanghai 200438, China}
\affiliation{Laboratory of Advanced Materials, Fudan University, Shanghai 200438, China}
\affiliation{Shanghai Research Center for Quantum Sciences, Shanghai 201315, China}

\author{Rui Peng}
\email{pengrui@fudan.edu.cn}
\affiliation{State Key Laboratory of Surface Physics, and Department of Physics, Fudan University, Shanghai 200438, China}
\affiliation{Laboratory of Advanced Materials, Fudan University, Shanghai 200438, China}
\affiliation{Shanghai Research Center for Quantum Sciences, Shanghai 201315, China}

\author{Donglai Feng}
\email{dlfeng@hfnl.cn}
\affiliation{New Cornerstone Science Laboratory, Hefei National Laboratory, Hefei, 230026, China}

\date{\today}

\begin{abstract}
Infinite-layer nickelates have been widely viewed as cuprate analogs in which superconductivity emerges and forms a superconducting dome centered around 10–20$\%$ cation substitution. Here we show that pristine and stoichiometric PrNiO$_2$, without cation substitution, exhibits intrinsic superconductivity characterized by zero resistance and diamagnetism in uncapped films. Through heterostructure engineering, we further exclude an interfacial origin of the superconductivity. Remarkably, zero-resistance superconductivity is consistently observed in trivalent-substituted PrNiO$_2$, whereas it is rapidly suppressed by dilute divalent substitution. Combined with angle-resolved photoemission studies, these results indicate that such a new superconducting regime is confined to within 3\% additional hole doping from pristine PrNiO$_2$. Furthermore, this phase is separated from the previously established superconducting dome around $\sim20\%$ divalent doping by a non-superconducting region in the phase diagram, and is further distinguished by a remarkably stronger upper-critical-field anisotropy. These findings establish a unique separated superconducting regime, suggesting that infinite-layer nickelates are not merely cuprate analogs but host distinct superconducting physics.
\end{abstract}

\maketitle



$Introduction-$The discovery of superconductivity in hole-doped infinite-layer nickelates~\cite{li2019superconductivity}, whose parent electronic configuration is nominally Ni 3$d^9$ as in cuprates, has generated tremendous interest in the search for cuprate analogs. So far, superconductivity has been established mainly in cation-substituted infinite-layer nickelates involving divalent or mixed-valence cations, including Sr, Ca and Eu~\cite{chow2025bulk,yang2025enhanced,osada2020phase,osada2020superconducting,wei2023superconducting,zeng2022superconductivity,osada2021nickelate,li2019superconductivity}. Accordingly, the phase diagram of infinite-layer nickelates has often been discussed within a cuprate-like framework, in which superconductivity only emerges upon cation doping and forms a dome around 10–20\% divalent-cation substitution~\cite{osada2020phase,osada2020superconducting,zeng2022superconductivity,li2020superconducting}. However, whether nickelate superconductivity is governed by a cuprate-like single dome remains unclear.

This question becomes particularly acute near the parent limit. In cuprates, the parent compounds are antiferromagnetic Mott insulators. By contrast, pristine RENiO$_2$ (RE = La, Pr, Nd, etc.) compounds are correlated metals with intrinsic self-hole doping from interstitial $s$ electrons~\cite{zhang2020self}. Recent angle-resolved photoemission spectroscopy (ARPES) experiments further show that the Ni $d_{x^2-y^2}$ orbital in pristine RENiO$_2$ already contains about 9\% holes~\cite{li2025observation,ding2024cuprate}, a doping level that would fall within the superconducting regime of hole-doped cuprates. Intriguingly, several recent studies have reported signatures suggestive of superconductivity near the parent limit~\cite{sahib2025superconductivity,parzyck2025superconductivity,osada2021nickelate,dong2025topochemical}. However, zero resistance has not been established in uncapped parent films~\cite{sahib2025superconductivity,parzyck2025superconductivity,osada2021nickelate,dong2025topochemical}, and diamagnetic response has not been reported, leaving open whether the reported signatures reflect intrinsic superconductivity of the pristine parent phase or arise from capping-layer-induced interfacial effects or unintended carrier doping.
Moreover, the key question lies  not only on whether intrinsic superconductivity can emerge near the parent limit, but also on whether it represents the low-doping edge of the established hole-doped dome or defines a distinct superconducting regime. Resolving this issue is essential for understanding the nickelate phase diagram and how superconductivity develops.

\begin{figure*}[t]
\includegraphics[width=170mm]{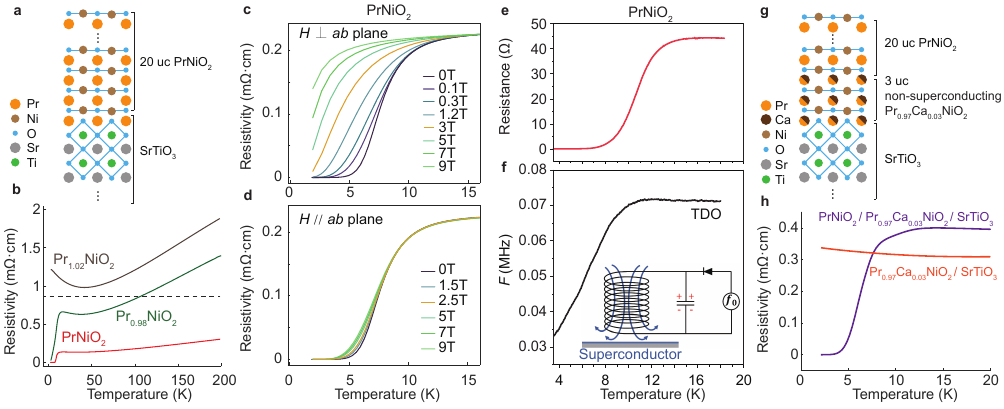}
\caption{\label{Figure1} 
\textbf{Intrinsic superconductivity in PrNiO$_2$/SrTiO$_3$ films.}
\textbf{a,} Schematic of 20~uc~PrNiO$_2$/SrTiO$_3$ films without SrTiO$_3$ capping layers.
\textbf{b,} Temperature-dependent resistivity of PrNiO$_2$/SrTiO$_3$ compared with off-stoichiometric samples.  The dashed line represents quantum of resistance per NiO$_2$ sheet $\rho_Q\sim=0.859~\rm{m}\Omega\cdot\rm{cm}$.
\textbf{c, d,} Magnetic-field-dependent transport of the PrNiO$_2$ thin films as a function of strong sweeping temperatures at various fixed magnetic fields, with the field oriented (\textbf{c}) in the $ab$ plane and (\textbf{d}) parallel to the $ab$ plane.
\textbf{e,} Low-temperature resistance of the PrNiO$_2$/SrTiO$_3$ film, highlighting the superconducting transition. 
\textbf{f,} Diamagnetic response of the PrNiO$_2$ film probed by a tunnel diode oscillator (TDO). The black solid line represents the TDO frequency shift, indicating diamagnetic response. The inset displays a schematic diagram of the TDO setup. 
\textbf{g,} Schematic of the 20~uc PrNiO$_2$/~3~uc Pr$_{0.97}$Ca$_{0.03}$NiO$_2$/SrTiO$_3$ heterostructure, where the 3~uc Pr$_{0.97}$Ca$_{0.03}$NiO$_2$ acts as a non-superconducting buffer layer to isolate interfacial charge transfer.
\textbf{h,} Temperature-dependent resistivity of the heterostructure compared with a bare Pr$_{0.97}$Ca$_{0.03}$NiO$_2$ films, confirming the intrinsic nature of the superconductivity in the upper PrNiO$_2$ layers.}
\end{figure*}

Here we show zero resistance and a clear diamagnetic response in high-quality, stoichiometric, uncapped PrNiO$_2$ films.  This superconductivity persists when the PrNiO$_2$ layer is separated from SrTiO$_3$, excluding an interfacial origin. These results establish superconductivity as an intrinsic property of pristine infinite-layer nickelates without divalent cation doping. 
Combined transport and ARPES measurements reveal a narrow superconducting regime near the parent state, separated from the established substantially hole-doped dome by an intervening non-superconducting doping range.
Our findings demonstrate that nickelate superconductivity is not captured by a simple cuprate-like doping paradigm.




$Intrinsic~superconductivity-$Figure 1a illustrates the heterostructure investigated: 20-unit-cell (uc) PrNiO$_2$/SrTiO$_3$ films without any capping layer (detailed growth procedures are provided in the Methods). To establish the intrinsic properties of uncapped PrNiO$_2$, we performed an extensive optimization of film stoichiometry. As shown in Fig.~1b, films with either a 2\% Pr excess or a 2\% Pr deficiency exhibit only partial resistance drops at low temperatures (see Supplementary Fig.~S2 for details). In contrast, when the Pr/Ni ratio is tuned to stoichiometry, the normal-state resistivity is substantially reduced, and the low-temperature upturn below 40~K is strongly suppressed (Fig.~1b).
The normal-state resistivity of stoichiometric PrNiO$_2$ films are well below the quantum of resistance per NiO$_2$ sheet, $\rho_Q$ (dashed line in Fig.~\ref{Figure1}b) across the entire temperature range, suggesting high quality with reduced Anderson localization~\cite{haviland1989onset,parzyck2025superconductivity,ji2026emergent}.
At lower temperatures, the stoichiometric film undergoes a superconducting transition and reaches a zero-resistance state (Fig.~\ref{Figure1}b), highlighting the critical role of stoichiometry and crystalline quality in realizing superconductivity in undoped PrNiO$_2$.

The transition to a zero-resistance state is observed across eight stoichiometric PrNiO$_2$ films (Supplementary Information Fig.~S1), demonstrating the reproducibility of superconductivity in this regime. Among all prepared films, the highest onset transition temperature, $T_{\mathrm{c}}^{90\%}$, is around 11~K, and a zero-resistance temperature, $T_{\mathrm{c0}}$ is around 7~K. The zero-resistance state is progressively suppressed by magnetic fields applied either parallel or perpendicular to the film plane (Figs.~\ref{Figure1}c,d), hallmark of superconducting behavior. Furthermore, we probed the diamagnetic response using a tunnel diode oscillator (TDO) circuit (see Methods and Supplementary Information Fig.~S3 for the experimental configuration), in which superconducting diamagnetic shielding of the sample induces a shift in the resonant frequency, $f_{\mathrm{TDO}}$. A clear frequency shift develops below $T_{\mathrm{c0}}$ (Fig.~\ref{Figure1}f), indicating diamagnetic shielding. Together with the observation of zero resistance, this establishes the two defining hallmarks of superconductivity in uncapped PrNiO$_2$. The absence of SrTiO$_3$ capping further rules out the top PrNiO$_2$/SrTiO$_3$ interface as the origin of superconductivity near zero doping.

To further test a possible substrate-interface origin of superconductivity, we inserted a 3-uc-thick non-superconducting Pr$_{0.97}$Ca$_{0.03}$NiO$_2$ spacer layer between PrNiO$_2$ and SrTiO$_3$, forming a PrNiO$_2$/Pr$_{0.97}$Ca$_{0.03}$NiO$_2$/SrTiO$_3$ heterostructure (Fig.~\ref{Figure1}g). This non-superconducting spacer layer separates PrNiO$_2$ from the SrTiO$_3$ substrate and is expected to strongly suppress substrate-induced chemical intermixing, charge transfer, and interfacial coupling. If superconductivity were primarily driven by such interface-related mechanisms, both the zero-resistance state and the transition temperature should be substantially weakened. However, the heterostructure retains a robust zero-resistance transition (Fig.~\ref{Figure1}h), with a $T_c$ comparable to that of stoichiometric PrNiO$_2$ films and falling within the sample-to-sample variation observed among the eight uncapped films (Supplementary Information Fig.~S1). The magnetic-field response of the resistivity is shown in Supplementary Information Fig.~S4. These results rule out a dominant role of the SrTiO$_3$ substrate interface and establish superconductivity in nominally undoped PrNiO$_2$ as intrinsic to the infinite-layer nickelate itself.

\begin{figure}[b]
\includegraphics[width=86mm]{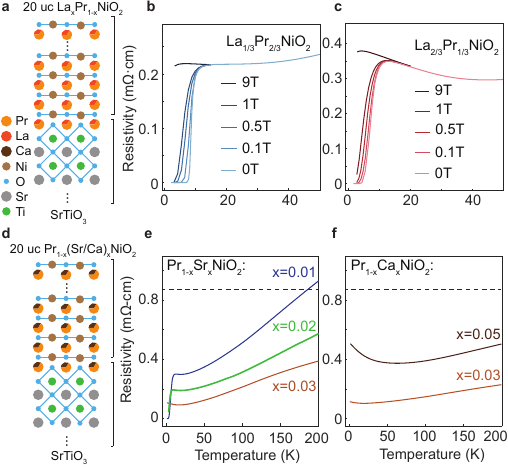}
\caption{\label{Figure2} \textbf{Superconducting robustness under isovalent substitution and its sensitivity to Sr/Ca doping.}
\textbf{a,} Schematic of 20~uc~La$_{x}$Pr$_{1-x}$NiO$_2$/SrTiO$_3$ films.
\textbf{b, c,} Magnetic-field-dependent transport of (\textbf{b}) La$_{1/3}$Pr$_{2/3}$NiO$_2$ and (\textbf{c}) La$_{2/3}$Pr$_{1/3}$NiO$_2$ under magnetic fields along $c$ axis, respectively. 
\textbf{d,} Schematic of 20~uc aliovalent-doped Pr$_{1-x}A_x$NiO$_2$ ($A$ = Sr, Ca) thin films. 
\textbf{e, f,} Temperature-dependent resistivity of (\textbf{e}) Pr$_{1-x}$Sr$_x$NiO$_2$/SrTiO$_3$ (x~=~0.01,~0.02 and 0.03) and (\textbf{f}) Pr$_{1-x}$Ca$_x$NiO$_2$/SrTiO$_3$ (x~=~0.03 and 0.05), respectively.  The dashed line represents quantum of resistance per NiO$_2$ sheet $\rho_Q$.
}
\end{figure}


$Contrasting~effects~of~isovalent~and~divalent~substitution-$The establishment of intrinsic superconductivity in uncapped PrNiO$_2$ raises the question of whether this state is specific to PrNiO$_2$ or reflects a more general feature of undoped infinite-layer nickelates. To address this question, we performed isovalent substitution of Pr$^{3+}$ by La$^{3+}$, which preserves the nominal carrier concentration while modifying the rare-earth environment (Fig.~\ref{Figure2}a). As shown in Figs.~\ref{Figure2}b,c, both La$_{1/3}$Pr$_{2/3}$NiO$_2$ and La$_{2/3}$Pr$_{1/3}$NiO$_2$ exhibit a pronounced superconducting transition and reach a zero-resistance state at low temperatures, demonstrating that superconductivity remains robust under substantial La substitution. Similar to pristine PrNiO$_2$, the superconducting state is systematically suppressed by magnetic fields applied (Figs.~\ref{Figure2}b,c). Notably, La$_{1/3}$Pr$_{2/3}$NiO$_2$ film exhibits a zero-resistance transition temperature of $T_{\mathrm{c0}}\sim7$ K comparable to PrNiO$_2$, suggesting that intrinsic superconductivity is not unique to PrNiO$_2$ and can emerge in well-optimized undoped infinite-layer nickelate films.


Superconductivity has previously been established in Pr$_{1-x}$Sr$_x$NiO$_2$ near $x\sim0.2$ (refs.~\citenum{osada2020phase,osada2020superconducting,wang2022pressure}), raising the question of whether the parent superconducting state evolves continuously into the hole-doped dome or represents a distinct superconducting phase. Dilute Sr substitution in PrNiO$_2$ provides a direct route to probe this connection (Fig.~\ref{Figure2}d).
As shown in Fig.~\ref{Figure2}e, 
the $x=0.01$ film still reaches a zero-resistance state near 2 K, while only a partial resistivity drop is observed for $x=0.02$, and the $x=0.03$ film exhibits weakly insulating behavior below 10 K without any indication of superconductivity. These results indicate that the superconducting phase near zero doping is confined within 3\% Sr doping.
To test whether the rapid suppression of superconductivity is specific to Sr substitution, we further examined aliovalent Ca substitution on the Pr site. Because the ionic radius of Ca$^{2+}$ is closer to that of Pr$^{3+}$, Ca substitution provides an alternative hole-doping route with potentially reduced lattice distortion~\cite{zeng2022superconductivity,chow2023dimensionality,hameed2025interplay,kim2021optical}.
As shown in Fig.~\ref{Figure2}f, Pr$_{1-x}$Ca$_x$NiO$_2$ films with $x=0.03$ and $0.05$ exhibit no signatures of superconductivity and instead display weakly insulating behavior at low temperatures. 
To further verify whether the absence of superconductivity in lightly hole-doped Pr$_{1-x}$Ca$_x$NiO$_2$ is intrinsic or a consequence of limited quality, we carried out systematic stoichiometry optimization on Pr$_{0.97}$Ca$_{0.03}$NiO$_2$ films. Despite improvements in crystallinity and normal-state transport properties, no superconducting transition was observed even in the optimally tuned films (see Supplementary Fig.~S5). These results confirm that the disappearance of superconductivity in the 3\% Ca doped PrNiO$_2$ does not arise from non-ideal stoichiometry or disorder.
Therefore, these results demonstrate that superconductivity is confined to a narrow doping range near the undoped limit, distinct from the established superconducting dome at higher hole concentrations.



\begin{figure*}[thb]
\includegraphics[width=170mm]{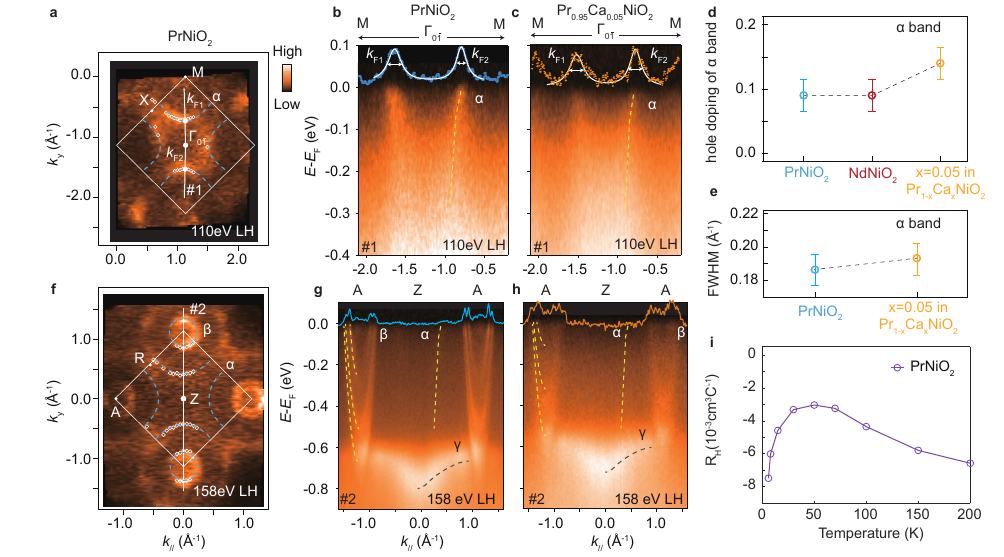}
\caption{\label{Figure3} \textbf{Comparable self-doping in PrNiO$_2$ with other pristine IL-nickelates and the suppression of superconductivity by slight hole doping.}
 \textbf{a,} Photoemission intensity map of PrNiO$_2$ at the $\Gamma$ plane, integrated over an energy window of $E_{\rm F} \pm 50$~meV under linear horizontal (LH) polarization. 
\textbf{b, c,} Band spectra of 20~uc PrNiO$_2$ (\textbf{b}) and Pr$_{0.95}$Ca$_{0.05}$NiO$_2$ (\textbf{c}) along the M-$\Gamma$-M direction. The momentum distribution curves (MDCs) integrated within $E_{\rm F} \pm 25$~meV are plotted above the respective spectra, where white arrows denote the full width at half maximum (FWHM) of the $\alpha$ band.
\textbf{d,} Effective hole-doping level of the Ni $3d_{x^2-y^2}$ ($\alpha$) band for PrNiO$_2$, NdNiO$_2$ and Pr$_{0.95}$Ca$_{0.05}$NiO$_2$ extracted from the Luttinger volume. 
\textbf{e,} Comparable full width at half maximum (FWHM) of the $\alpha$ band in PrNiO$_2$ and Pr$_{0.95}$Ca$_{0.05}$NiO$_2$.
\textbf{f,} Photoemission intensity map of PrNiO$_2$ at the Z plane ($E_{\rm F} \pm 50$~meV, LH polarization). 
\textbf{g, h, } Band spectra of 20~uc PrNiO$_2$ (\textbf{g}) and Pr$_{0.95}$Ca$_{0.05}$NiO$_2$ (\textbf{h}) along the A-Z-A direction, with corresponding MDCs ($E_{\rm F} \pm 25$~meV) shown above. 
\textbf{i,} Hall coefficient R$_{\rm{H}}$ of superconducting PrNiO$_2$ versus $T$. }
\end{figure*}

\begin{figure*}[htb]
\includegraphics[width=170mm]{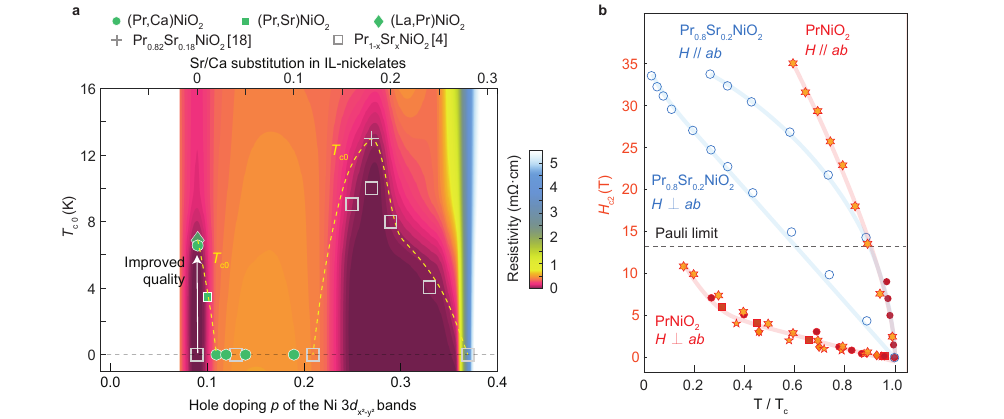}
\caption{\label{Figure4} \textbf{Distinct superconducting regime from optimally hole-doped superconducting dome with more two-dimensional superconductivity.} 
\textbf{a}, Phase diagram of hole doped PrNiO$_2$ and equivalent cationic rare-earth ion substituted La$_{x}$Pr$_{1-x}$NiO$_2$. The hole doping $p$ of the Ni $3d_{x^2-y^2}$ band was determined by the Luttinger volume of uncapped IL-nickelates, Pr$_{0.95}$Ca$_{0.05}$NiO$_2$ and Pr$_{0.8}$Ca$_{0.2}$NiO$_2$ according to ARPES results from this work and refs.~\citenum{ding2024cuprate,li2025observation}, respectively. Resistivity data of Pr$_{1-x}$Sr$_{x}$NiO$_2$ around the optimally-doped regime is adapted from refs.~\citenum{osada2020phase,wang2022pressure}, respectively.
\textbf{b,} Upper critical field $H_{\rm{c2}}$ of PrNiO$_2$ samples versus $T/T_{\rm{c}}$. The hollow circles refer to data of optimally doped Pr$_{1-x}$Sr$_{x}$NiO$_2$, which is adapted from ref.~\citenum{wang2023effects}. The black dashed line represents the Pauli limit of PrNiO$_2$ samples $H_{\rm{P}}=1.86\times T_{\rm{c,50\%}}=14.18~\rm{T}$. 
}
\end{figure*}

$Electronic~structure-$To understand the band structure related with the superconducting phase around $x$~=~0, we performed \textit{in-situ} ARPES measurements on bare superconducting PrNiO$_2$ and non-superconducting Pr$_{0.95}$Ca$_{0.05}$NiO$_2$ films (Fig.~\ref{Figure3}). Overall, the Fermi surfaces of PrNiO$_2$ (Fig.~\ref{Figure3}a,f) consist of a hole-like Ni $3d_{x^2-y^2}$-derived $\alpha$ pocket both around M at the $\Gamma$-X-M plane and around A in the Z-R-A plane, suggesting the quasi-two-dimensional character of the $\alpha$ pocket. There is a $\beta$ pocket around A point at Z-R-A plane. This Fermi surface topology closely resembles that of apical-oxygen-free NdNiO$_2$ films~\cite{li2025observation}, and quantitative analysis gives a comparable hole concentration of approximately $p=0.09$ in the Ni $3d_{x^2-y^2}$ orbital, consistent with the self-doped parent electronic structure of infinite-layer nickelates (see detailed comparison in Supplementary Fig.~S7).
To further examine whether superconducting PrNiO$_2$ remains in the parent self-doped regime, Hall measurements were performed. As shown in Fig.~\ref{Figure3}i, PrNiO$_2$ exhibits a negative Hall coefficient over the entire measured temperature range, closely resembling the behavior of NdNiO$_2$ (ref.~\citenum{zhang2025thermoelectricity}). This behavior contrasts with the superconducting films in the established hole-doped dome around 20\% Sr or Ca, which display a positive Hall coefficient throughout the temperature range \cite{li2020superconducting,osada2020phase,osada2021nickelate,zeng2022superconductivity}. 
The negative Hall response reflects the multiband nature of PrNiO$_2$, in which the electron-like $\beta$ pocket near the A point  (Figs.~\ref{Figure3}g,h), together with the hole-like $\alpha$ pocket, contributes to the Hall coefficient.
Together, the ARPES and Hall measurements demonstrate that superconducting PrNiO$_2$ retains the characteristic electronic structure and carrier density of the undoped parent state, excluding unintended hole doping as the origin of superconductivity.

The evolution of this parent electronic structure upon Ca doping was examined using non-superconducting Pr$_{0.95}$Ca$_{0.05}$NiO$_2$ films.
As shown in Figs.~\ref{Figure3}c,h, the $\alpha$ and $\beta$ bands remain well defined and continue to cross $E_{\rm F}$, indicating that the overall band topology is preserved upon Ca substitution. 
The $\gamma$ band shifts towards $E_{\rm F}$ (Figs.~\ref{Figure3}g,h), and the momentum distribution curves (MDCs)  integrated within $E_{\rm F}\pm25$ meV reveal a reduced separation between the two Fermi crossings, $k_{\rm F1}$ and $k_{\rm F2}$, of the $\alpha$ band (Figs.~\ref{Figure3}b,c). This change corresponds to an enlarged $\alpha$ hole pocket, indicating that Ca substitution effectively introduces additional holes into the NiO$_2$ planes (Fig.~\ref{Figure3}d).
Note that PrNiO$_2$ and Pr$_{0.95}$Ca$_{0.05}$NiO$_2$ exhibit comparable MDC linewidths for the $\alpha$ band (Figs.~\ref{Figure3}b,c,e, and see detailed comparison in Supplementary Fig.~S8), demonstrating that disorder alone is unlikely to account for the rapid suppression~\cite{sobota2021angle}. The observation of multiple Fermi crossings associated with the $\beta$ band (Supplementary Fig.~S9), consistent with quantum-well states~\cite{li2025observation}, further indicates highly homogeneous electronic states throughout the film thickness. Moreover, recent studies on (Sm,Eu,Ca)NiO$_2$ films have shown that Ca incorporation can preserve, and even improve, the structural quality and superconducting properties of infinite-layer nickelate films~\cite{chow2025bulk,yang2025enhanced}. This is compatible with the negligible MDC broadening observed upon Ca substitution~(Figs.~\ref{Figure3}b,c,e, and Supplementary Fig.~S9), suggesting that any additional disorder-induced scattering remains limited.
These observations suggest that Ca substitution effectively dopes additional holes into the NiO$_2$ planes, without introducing additional disorder. The rapid disappearance of superconductivity is not likely due to impurity scattering, but instead reflects the suppressed superconductivity upon small hole doping. 
$A~distinct~superconducting~regime-$Figure~\ref{Figure4}a summarizes these results in an effective hole-doping phase diagram for infinite-layer nickelates, using the Ni $3d_{x^2-y^2}$ hole concentration estimated from ARPES as a reference scale for the parent and Ca-substituted films. Superconducting PrNiO$_2$ and (La,Pr)NiO$_2$ films lie near the self-doped parent value of $p\approx0.09$, where zero resistance emerges with $T_{\mathrm{c0}}$ reaching $\sim7$ K. With dilute Sr or Ca substitution, superconductivity is rapidly suppressed and vanishes within only 3\% divalent-cation substitution, corresponding to a small increase in effective hole doping. 
Films with 5\% and 10\% Ca substitution also remain non-superconducting and exhibit weakly insulating behavior at low temperatures (Supplementary Fig.~S6). Superconductivity re-emerges at approximately 15\% Sr substitution and evolves into the established hole-doped dome at higher doping levels~\cite{osada2020phase,wang2022pressure}. These results establish that the narrow superconducting regime near the parent state is disconnected from the conventional hole-doped dome.

Furthermore, high-field transport measurements reveal that the parent superconducting regime exhibits an upper-critical-field response different from the established hole-doped dome (Fig.~\ref{Figure4}b).
For fields applied parallel to the NiO$_2$ planes, $H_{\mathrm{c2}}$ of PrNiO$_2$ reaches 35 T already at $T/T_{\mathrm{c}}\sim0.6$, the highest field accessible in our measurements, well above the Pauli paramagnetic limit. By contrast, the out-of-plane $H_{\mathrm{c2}}$ is much smaller and shows a low-temperature upturn, possibly reflecting Pauli-paramagnetic pair-breaking effects~\cite{wei2023large}. As a result, the upper-critical-field anisotropy, defined as $\gamma = H^{ab}_{\mathrm{c2,50\%}}/H^{c}_{\mathrm{c2,50\%}}$, reaches approximately 68 near $T_{\mathrm{c,50\%}}$, substantially exceeding values reported for optimally doped nickelates~\cite{sun2023evidence,wei2023large,osada2025systematic,wang2023effects,wang2021isotropic} (see Supplementary Table~1).
Using the high-field data within the Ginzburg–Landau framework~\cite{wang2021isotropic,sun2023evidence,harper1968mixed}, the in-plane coherence length is extracted to be $\xi_{ab}(0)=6.23$~nm, larger than that reported for optimally doped IL-nickelates ($\xi_{ab}(0)=2-4$ ~nm)~\cite{osada2021nickelate,zeng2022superconductivity,sun2023evidence,wang2021isotropic}.  
The analysis further yields an effective superconducting layer thickness of $d_{\rm sc}=3.56$ nm, corresponding to more than ten unit cells and within the same order of magnitude as the 6.6-nm film thickness. Although this estimate is not a precise microscopic thickness, it supports an extended superconducting response, consistent with the uncapped and interface-engineered heterostructure experiments (Fig.~1) against an interface-dominated origin.
The enhanced anisotropy provides an additional phenomenological distinction between the parent superconducting regime and the substantially hole-doped regime.

$Discussion-$Once stoichiometry and crystalline quality are sufficiently optimized, pristine (La,Pr)NiO$_2$ films can host superconductivity without additional carrier doping. At first sight, this may appear consistent with a cuprate-like doping picture, because ARPES shows that nominally undoped PrNiO$_2$ already contains approximately 9\% holes in the Ni $3d_{x^2-y^2}$ orbital, placing it in a filling range comparable to underdoped cuprates within the superconducting dome~\cite{zhang2020self,ding2024cuprate,keimer2015quantum}. 
However, the doping evolution observed here is difficult to reconcile with such a simple picture. If the parent superconducting phase were merely the low-doping edge of a conventional hole-doped dome, increasing the effective hole concentration from  9~\% to about 12~\% would be expected to move the system closer to optimal doping and to  enhance superconductivity. Instead, superconductivity is rapidly suppressed. One might consider an analogy to the stripe-related suppression of superconductivity near the $1/8$ anomaly in cuprates, where competing electronic orders can weaken superconductivity at particular dopings. However, in PrNiO$_2$, superconductivity is absent over a broad doping range between the parent superconducting regime and the substantially hole-doped dome, rather than being suppressed only near a particular doping associated with stripe order~\cite{tranquada1995evidence,abbamonte2005spatially,wu2011magnetic,keimer2015quantum}. In addition, our ARPES measurements on Pr$_{0.95}$Ca$_{0.05}$NiO$_2$ show no clear band folding or Fermi-surface reconstruction, suggesting that long-range stripe order is unlikely. 
These results show that superconductivity near the parent limit cannot be understood simply as the low-doping extension of a cuprate-like superconducting dome.


The disconnected phase diagram, together with the markedly enhanced upper-critical-field anisotropy, points to a superconducting state that may differ qualitatively from that realized in substantially doped regime. In this context, theoretical proposals of unconventional $(d+is)$ pairing accompanied by time-reversal-symmetry breaking provide one possible route for understanding the distinct properties of the undoped superconducting phase~\cite{wang2020distinct}. 
Another possibility is that superconductivity near the parent state is connected to the boundary of a putative electron-doped superconducting regime. 
Recent calculations suggest that electron doping enlarges rare-earth/interstitial-$s$ derived electron pockets and preserves robust antiferromagnetic interactions, giving rise to a regime distinct from the substantially hole-doped superconducting dome~\cite{dayroberts2026electron}. 
In this scenario, small hole doping would naturally drive the system away from the electron-pocket-dominated superconducting instability, thereby suppressing superconductivity before the conventional hole-doped dome is reached.

Overall, the establishment of an intrinsic superconducting phase near the undoped limit reshapes the superconducting phase diagram of infinite-layer nickelates. The presence of two separated superconducting regimes, one emerging from the self-doped parent state and the other from substantial hole doping, places strong constraints on microscopic theories and calls for a reassessment of how electronic correlations, competing orders, and pairing symmetry evolve across the novel phase diagram. These findings define a new framework for understanding infinite-layer nickelate superconductivity and its relationship to the cuprate paradigm.

\nolinenumbers

$Acknowledgments-$We acknowledge F. C. Zhang, Y. J. Yu, S. Y. Li, L. Shu, X. Zhang, and Y. Wang for helpful discussions. This work was supported by the National Natural Science Foundation of China (Grants Nos. 92477206, 12422404, 12274085, 12474053), the National Key R\&D Program of China (Grant Nos. 2023YFA1406300, 2022YFA1602602, 2024YFF0727900), the New Cornerstone Science Foundation, the Innovation Program for Quantum Science and Technology (Grant No. 2021ZD0302803), the Shanghai Municipal Science and Technology Major Project (Grant No. 2019SHZDZX01). We thank the Shanghai Synchrotron Radiation Facility (SSRF) BL03U (31124.02.SSRF.BL03U) for assistance with ARPES measurements and the Water-Cooled Magnet 5 (WM5) at the Steady High Magnetic Field Facility, CAS (31125.02.SHMFF.WM5), for technical support in magnetic-field-dependent transport measurements and data analysis.




\appendix

\bibliography{apssamp}
\end{document}